# Barry Carpenter – Pioneering Physical Organic Chemist, Teacher, Mentor and Friend

Stephanie R. Hare,[¤] Allan R. Pinhas,[#] Julia Rehbein,[&] Dean J. Tantillo,[‡] and Stephen Wiggins[§]

[¤] *Atomwise, 717 Market St #800, San Francisco, CA 94103*

[#] *Department of Chemistry, University of Cincinnati, Cincinnati, OH, USA*

[&] *Institute of Organic Chemistry, University of Regensburg, Universitätsstr. 31, 93053 Regensburg, Germany*

[‡] *Department of Chemistry, University of California–Davis, Davis, CA 95616 USA*

[§] *School of Mathematics, University of Bristol, Fry Building, Woodland Road, Bristol, BS8 1UG, United Kingdom*

This special issue of the *Journal of Physical Organic Chemistry* is dedicated to an outstanding physical organic chemist, mentor and friend, Barry Carpenter on the occasion of his official retirement. Knowing that great minds like his never fully retire, we thought a fitting tribute would be a compilation of recent research by people who have been lucky enough to have crossed paths – trajectories – with him! Here you will find a selection of contributions from Barry's science family. We hope that the diversity of these contributions will encourage readers to broaden their views on reactivity and to delve deeper into its origins – as Barry's many contributions, in papers, talks, and personal discussions, have done!

To Barry, we say: Thank you for having opened doors in our minds, for leading us beyond the well-trodden minimum energy path and into new adventures of chemistry!

Below we include: (1) an autobiographical sketch provided by Barry himself. (2) Testimonials from colleagues and friends speaking to Barry's contributions both professional and personal. (3) A comprehensive list of Barry's publications.

**Barry in Barry's own words**

I was born in Hastings, in SE England, and lived there until 1967, when I left to go to college. As a child, I had become interested in chemistry and biochemistry after discussions with my older brother, who was the first in my family to pursue higher education, and who had studied biochemistry. My interest was further piqued by being given a chemistry set as a gift. It contained substances, such as sodium dichromate, that would never be permitted to be sold to children today. Even more remarkably, it was perfectly acceptable for a 12-year-old to go to the local pharmacy with a glass-stoppered bottle and have it filled with concentrated hydrochloric or even nitric acid! I am probably lucky not to have killed myself in my garden-shed laboratory. On the other hand, I am not sure whether the tepid items that pass for chemistry sets these days would have stimulated me to pursue chemistry as a career.

My higher education began at the University of Warwick, which had only opened two years before my arrival. It had a real pioneering spirit, although was also clearly still a work in progress. Rather than having hallowed, ivy-covered buildings of centuries duration, it consisted of blocks of white-tiled structures, causing some wits to liken it to a giant public toilet! The principal hazard of being a student there at the time was avoiding the tiles that would fall off the buildings with some regularity. In line with the general Zeitgeist of the late 1960s, there was a general wish to try new directions in education at Warwick. Germaine Greer was a lecturer in the English department at the time, for example. In the sciences, there was a similar wish to explore new ideas. So, my degree course was actually in Molecular Sciences, rather than chemistry. Instead of teaching chemistry in a conventional fashion with what one might call the traditional, vertical divisions of organic, inorganic, and physical chemistry, there was an effort to take a topic, say stereochemistry, and to explore its meaning and importance across the traditional disciplines. Thus, the general field of chemistry was divided horizontally rather than vertically. I believe my exposure to this way of thinking influenced by own rather generalist tendencies later in life.

I had originally thought that I would follow my brother's path and pursue biochemistry, but at Warwick I was inspired by one particular lecturer, John M. Brown, who showed me the delightful logic of physical organic chemistry. I decided that that was what I really wanted to do, and so pursued a doctorate in the topic with Prof. H. M. R. Hoffmann at University College London. I worked on the chemistry of "oxyallyl" cations (the class of reactive intermediate formally formed by breaking the 2,3 C–C bond of a cyclopropanone). I also dabbled in a bit of organometallic chemistry, exploring the question of whether the newly developed Woodward-Hoffmann rules applied to organometallic as well as organic reactions.

Following my receipt of a Ph.D. in 1973, I headed off to the USA for postdoctoral studies. I applied to several of the big-name American physical organic chemists of the time, but was accepted by only one of them – Jerome Berson at Yale University. As it turned out, this was a very fortunate outcome, because I received the best possible education in Berson's lab. I felt at the time, and still believe today, that I learned more in the 18 months with Berson than the 6 years of chemistry education that I had had to that point. Berson was an outstanding mentor. He was not prescriptive about what one should study, but was absolutely rigorous about how one should study it. His outstanding intellect combined, I must say, with his qualities as just a thoroughly decent human being, had a very strong impact on me. With hindsight, it is no surprise at all that many of his student and postdocs went on to become outstanding independent researchers – sometimes in fields far removed from traditional physical organic chemistry.

I had developed an interest in computational chemistry during my Ph.D. and that interest grew during my time with Berson. However, doing computer calculations of any kind was a bit different then from the way it is today.  I had to carry heavy boxes of punched cards across the street to the Yale computing center (and heaven forbid that one should slip on an icy path in the wintertime and drop the box!).  The cards were left there and after, typically, a couple of days one could pick them up again, along with the output of the calculations on massive, folded sheets of paper.  This exercise did not encourage one to try much in the way of exploratory calculation!

As the time for my postdoctoral study drew to a close, the question arose about what I would do next.  I had hoped to return to a position as lecturer in the UK, but in the mid 1970s, the UK economy was in recession and there were essentially no positions available.  I consequently looked for places in the US and was lucky enough to land an Assistant Professor job at Cornell University.  After just a few months at Cornell, it became clear to me that this was what I should have been doing all along.  The Department was very welcoming and there were strong social as well as scientific interactions among the faculty members.  I subsequently discovered that this was far from the norm among US academic departments, and so, just as I had done when I happened to land in Berson's lab, I had fallen into something of a bed of roses by nothing but blind luck.

The fact that I spent 31 years on the faculty at Cornell says everything necessary about how I felt about the place.  When I finally left in 2006, it was certainly not because I had developed any antipathy to the Department or to my friends and colleagues there.  It was that Cardiff University, back in the UK, had created a new Physical Organic Chemistry Centre, and was looking for a director.  They offered me the position, and I accepted it.  That is where I finished my career, and Wales is where I still live.

Over the many years of my career, I have been blessed to work with a large number of highly talented graduate students and postdocs.  They have taught me at least as much as I taught them, and so I take this opportunity to express my sincere gratitude for the many ways in which they enriched my life.

**Personal Testimonials (in alphabetical order by author last name)**

*From Craig Bayse*

I may never understand why Barry offered a postdoc to a computational inorganic chemist with no synthetic organic experience. Nevertheless, I have always appreciated how the opportunity altered my chemical thinking – arrow pushing inorganic chemists are a rarity after all! The experience and humility gained from working with Barry and my labmates over just two years in Ithaca made a huge impact on how I have approached research problems and student mentoring. I am grateful for Barry's guidance, patience, and generosity. Happy retirement, Barry!

*From Matthew Cremeens*

After taking one of Barry's graduate courses, I really wanted to see what he was like when teaching an undergraduate organic chemistry course. One day, I had the privilege of witnessing him lecture to largely 2nd year Cornell University undergraduates. During the lecture, he asked the auditorium full of students a question. I sat there wondering; I could not imagine how this would unfold. Seared into my memory was Barry waiting, very patiently waiting quite some time for a student to respond. Of course a student responded, but to me the real story was the sheer magnitude of his gentle patience for his students. Barry's persistent patience was witnessed during every single Carpenter research group meeting; he modeled for all of his students, whether they were his graduate students or undergraduates, that persistence and patience were key to teaching and learning.

*From Stephanie R. Hare*

From the beginning of graduate school, I considered Barry's papers essential reference material for physical organic chemistry (reference material that reads like poetry). It would not be hyperbolic to say that the vast majority of my understanding of kinetic rate theories and non-statistical dynamic effects came directly from his papers. Considering the applicability of his work to my own, it felt completely natural to pursue a postdoc position in his group. Though moving to a new country (US → UK) was a big life decision, with the Chemistry and Mathematics in Phase Space (CHAMPS) program starting right before I was planning to apply for postdoc positions, and considering the impact I knew his mentoring would have on my career, working for Barry was a pretty easy choice to make.

Working on our first paper together taught me everything I needed to know about writing research papers: push past uncertainty to get words on the page, prioritize scientific rigor and clarity above all else, and don't be precious about keeping writing or figures on which you have spent a lot of time. That last lesson was a huge obstacle for me to overcome, but Barry helped me realize that scientific writing is for others and not yourself, and thus you have to listen if the reader has critiques of how information is coming across.

When I made my unceremonious departure from the UK in September of 2020, due to the COVID-19 pandemic, I never got the chance to say goodbye to Barry in person. I am so thankful for everything Barry has done to contribute to training, teaching, and advancing not just my career, but the physical organic chemistry community as a whole – a community that feels like a sprawling scientific family. Happy retirement Barry!

*From Ken Houk*

Barry Carpenter has influenced me throughout his career. His early papers with his then-student Cindy Burrows on the effect of cyano substituents (at all five positions!) on the rate of the Claisen rearrangement interested me immensely (C. J. Burrows and B. K. Carpenter, *J. Am. Chem. Soc.* **1981**, *103*, 6983, 6984). These papers contained synthesis, kinetics, and theory – the essence of modern physical organic chemistry, done in his early career at Cornell. Nine years later, Wes Borden, Keiji Morokuma, and I studied cyano group effects on the Cope, inspired by the Burrows-Carpenter work.

I had many occasions to meet Barry after that at meetings, to learn from him, and to learn from his 1984 book, *Determination of Organic Reactions Mechanisms*, which I used for teaching, too. I enjoy immensely his urbane manner and delightful wit. I admired and relished his brave entry into a new world for physical organic chemistry – computational molecular dynamics. After he pioneered that field and showed the rich insights about mechanisms that were possible, many of us – Dan Singleton, Chuck Doubleday, and I, to name a few – followed him into this field. Now a host of other computational organic chemists have joined him to study time-resolved mechanisms and non-statistical views of how reactions occur.

His move to Cardiff fifteen years ago was a loss to U.S. science, but Barry continues to enrich the chemical community with his profound insights into how molecular reactions occur.

*From Thomas Hughes*

It goes without saying that Barry's contributions to science can be described as so elegant, aesthetically pleasing, and obviously true that you might be fooled into believing that you could have arrived at the same penetrating conclusion if you had just thought about the problem hard and long enough. You wouldn't have, of course. While it's certainly true that his insights have been penetrating, it's also his skill as a communicator and teacher that makes these insights so accessible. During my time in his research group, observing Barry's approach to teaching and his skills as a communicator has had a profound impact on me as a chemist and as an educator. What impressed me most wasn't his writing or pedagogy, as well-crafted and meticulous as they were, but rather his respect for the process of communication itself, and for the people engaged with him in that communication.

He recognizes the common humanity in each and every person who is sincerely seeking knowledge and understanding. Barry always showed the same patience when talking to an undergraduate student, one of his graduate advisees, or a colleague. I repeatedly saw him answer what could be recognized as a "dumb" question with the same outward respect and patience that he would show a Nobel laureate. It's even possible that I might have asked one of those questions. Now, in my career as an educator, I do everything I can every day to treat my students with this same respect. Barry showed me how fortunate we are to not only be chemists solving problems, but also to be able to share our solutions with other chemists and students. I know I've been fortunate to have Barry share some of those elegant and obviously true ideas with me. Thanks, Barry.

*From Jeehiun Lee*

It is my pleasure to relay my experiences with Barry Carpenter. I first met Barry when I took his famous graduate physical organic class at Cornell, Chemistry 665. (This course was followed the next semester by Chemistry 666, and I think I am probably not the only student who thought that Barry's challenging course should have that diabolical number!). I was an undergraduate at the time, doing synthetic chemistry and planning a career in that area. I still remember the first lecture – Barry trotted out the Schrödinger equation and some complex derivation that one does not expect in an organic course. But Barry of course had a plan – this was his lead-in to molecular orbital theory. I also remember the first exam probably had a mean of about 20 percent. None of this mattered to us, however, because Barry was such an exceptional teacher. He made the material come alive with his passion; but just as importantly, Barry has an amazing ability to clarify difficult concepts. The class was so inspiring that it made me fall in love with mechanistic organic chemistry and choose that as my career. Thus, I have Barry to thank for my entire career path -- and I am ever so grateful to him for that.

There is also a coda to this story. Many graduate students were so impressed by Barry's teaching that they recommended the course to me (since I was an undergraduate). Quite frankly, Chemistry 665 was rather legendary. Usually students complain about a hard course, but Barry was the kind of teacher who pushed one to rise to the challenge such that despite the course's difficulty, students sang the praises of both the class and Barry. One of those graduate students (Blair Wood) later became my husband, so I suppose you could say that I have Barry to thank for my entire personal life as well!

*From Allan Pinhas*

I first met Barry in September of 1975 and became the first member of his research group the following January.  Whether in the laboratory or in a classroom, education of students has always been paramount for Barry.  He helps his students grow as scientists and as individuals.  Barry was readily willing to answer any questions I had, no matter how intelligent or how trivial.  For my research project, while discussing what experiments were needed to understand the mechanism or to perform the synthesis, Barry never told me what to do; but rather he would ask, what reaction do you think should be done?  In general, he makes suggestions and helps guide his students, so they grew intellectually in this process.  If I went off in a different direction from what we discussed, Barry would inquire about the results and discuss them with me.  I have tried to pattern my own behavior with my students after Barry, and if I have been successful, then it is a result of my time in his research group.  Therefore, I would like to thank Barry for my growth as a scientist and as a person.

*From Ashoka Samuelson*

I became a part of the Carpenter group in the fall of 1978. Everyone in the group called him Barry. Coming from India where we addressed all our Professors as "Sir" or "Professor"  I could not call him Barry and mostly called him Prof. Carpenter. What impressed me about Barry was his approach to solving problems in mechanistic organic chemistry. His research problems were elegant. He was to mechanistic organic chemistry what Woodward was to synthetic organic chemistry, a "supreme patterner of chaos".

The freedom Barry gave his students was quite amazing. I wanted to take a course on experimental methods with Jim Burlitch in my second semester. Although it cut into my research time, he agreed.  He was easygoing with my request to work partly with him and partly with Roald. I did not know that he ran computations himself. He could have guided me on both fronts but he gave me the freedom to do what I wanted. Halfway through the semester, I could not handle attending two group meetings. I liked experimental work more and so I ditched Roald (he has remained a lifelong friend) and joined the Carpenter group fully. We had weekly problem-solving sessions in the Carpenter group which was the high point of the week. He was literally one among us and treated everyone with kindness. He hardly asked for work reports.

Barry's classes were well thought out and logically arranged. His lectures were very clear and we sat mesmerised. One would not have any doubts and his "chalk work" on the blackboard was very neat with structures drawn as good as one can draw with ChemDraw! It was only later that I realised that he must have spent hours planning those courses, writing those assignments and hand-outs and some of them were hand-written and photocopied.

Barry was also a great counsellor and comforter. One day I found one of the students literally seated on the floor and listening to the Guru! I was heartbroken when the problem I was trying to solve was published. Not Barry, he just moved on. Frankly, if this had happened to one of my students, I would have expected the student to work on a different problem to have a more "successful thesis". Barry only looked at the effort put in and let me graduate at the end of four years. Barry's mentoring was one of the best things that has happened in my life!

*From Dean J. Tantillo*

When I was a postdoc at Cornell, I attended Barry's group meetings in addition to those of my postdoc advisor, Roald Hoffmann.  I wanted to make sure I had enough mechanistic organic chemistry in my life! One of my lasting memories from this time is hearing Barry talk about non-statistical dynamic effects on organic reactivity. I use the correct words here, but I didn't understand the concept back then. Much later in my independent career I finally understood what non-statistical dynamic effects are and why they are important, and then realized that Barry knew all that decades ago. He was so far ahead of his time; nowadays, including such effects when modeling organic reactivity is becoming commonplace.  My time at Cornell led to one paper with Barry, on Fe-templated ring-opening of methylenecyclopropanes, but I wish I had the wherewithal to have grasped the importance of dynamic effects back then! Since that time, I have had the pleasure of interacting with Barry at various conferences and getting to know him a bit. I appreciate his opinions on science and politics (he was ahead of his time with respect to the latter as well). I have always been amazed at how he can find fatal flaws in approaches to scientific projects and reveal them to practitioners in helpful rather than hurtful ways. When I submit my best papers I always request Barry as a referee and tell my students that if we can get this work past Barry, it must be okay. I am grateful for the support Barry has given me and the mechanistic organic community, both directly and indirectly. He is a quiet giant and an inspiration.

*From Veronica Vaida, Elizabeth Griffith, Becky Rapf, Allison Reed-Harris, Russell Perkins*

My group's collaboration with Barry Carpenter had a memorable beginning. We had essentially given up on understanding the photolysis of aqueous pyruvic acid. In chatting with Barry, who was visiting the department at the time, we mentioned the particularly noticeable smell produced by this photochemistry. Barry came by the lab and literally "sniffed out" the mechanism that was producing acetoin. "That moment when Barry walked into the lab and proclaimed that he knew what the smell was, is one of my favorite memories from my time in your group" (Elizabeth).

This was the start of a wonderful, educational and fun collaboration with my group of physical chemists. Meetings with Barry around the table with the group, with Anne Monod and the NMR wizardry of Rich Shoemaker led some years and many studies later, to our appreciation of organic reaction mechanisms. Barry has always been wonderfully supportive and encouraging of our attempts at mechanism development, even when they were naïve or wrong. Through our collaborations, he has very patiently taught my group how to think chemically. Barry Carpenter was and still is a wonderful teacher and mentor. Working with him has always been fun, and still is.

**Barry's Publications to Date**

*Book:*
"Determination of Organic Reaction Mechanisms", Wiley-Interscience, New York, 1984.

*Book Chapters:*
"Preparation and Uses of Isotopically Labeled Derivatives," in "The Chemistry of the Cyclopropyl Group," S. Patai and Z. Rappoport, Eds., Wiley, Chichester, 1987

"Theory and Experiment in the Analysis of Reaction Mechanisms," in "Advances in Molecular Modeling", D. Liotta, Ed., JAI Press, Greenwich, CT, 1988.

"Electron Capture Dissocation Produces Many More Protein Backbone Cleavages than Collisional or *IR* Excitation," Zubarev, R.A.; Fridriksson, E.K.; Horn, D.M.; Kelleher, N.L.; Kruger, N.A.; Lewis, M.A.; Carpenter, B.K.; McLafferty, F.W. in "Mass Spectrometry in Biology and Medicine," Burlingame, A.L.; Carr, S.A.; Baldwin, M.A., Eds., Humana Press, Totowa, NJ, 1999.

"Potential Energy Surfaces and Reaction Dynamics," in "Reactive Intermediate Chemistry," R. A.Moss, M. S. Platz, and M. Jones, Eds., Wiley-Interscience; New York, 2004.

"Bicyclo[2.1.0]pentanes and Bicyclo[2.2.0]hexanes," in "The Chemistry of Cyclobutanes," Z. Rappoport, Ed., Wiley, Chichester, 2005.

*Papers*

(1) "Synthesis of 3(2H) Furanones," Carpenter, B.K.; Clemens, K.E.; Schmidt, E.A.; Hoffmann, H.M.R. *J. Am. Chem. Soc*. **1972**, *94*, 6213.

(2) "An INDO Molecular Orbital Study of 1,3 [π] Interactions in 2-Substituted Allyl Cations," Carpenter, B.K. *J. Chem. Soc. Perkin II*, **1974**, 1.

(3) "Synchronous Double Rotation in the Stereomutation of Optically Active 1-Phenylcyclopropane-2-$d_1$," Pedersen, L.D.; Berson, J.A.; Carpenter, B.K.*J. Am. Chem. Soc.* **1975**, *97*, 240.

(4) "Octamethyl-1,4-cyclohexanedione," Carpenter, B.K.; Rawson, D.I.; Hoffmann, H.M.R. *Chem. Ind. (London)* **1975**, 886.

(5) "Thermal Stereomutation of Cyclopropanes," Pedersen, L.D.; Berson, J.A.; Carpenter, B.K. *J. Am. Chem. Soc*. **1976**, *98*, 122.

(6) "The Influence of Substituents on the Molecular Orbital Energies and Ground Electronic States of Substituted Trimethylenemethanes," Carpenter, B.K.; Little, R.D.; Berson, J.A. *J. Am. Chem. Soc.* **1976**, *98*, 5723.

(7) "2,3-Diazabicyclo[2.2.0]hex-2-ene," Wildi, E.A.; Carpenter, B.K. *Tetrahedron Lett*. **1978**, 2469.

(8) "Frontier Orbital Control of Regiospecificity in Singlet Cycloadditions of 2-Methylenecyclopenta-1,3-diyls," Siemionko, R.; Shaw, A.; O'Connell, G.; Little, R.D.; Carpenter, B.K.; Shen, L.; Berson, J.A. *Tetrahedron Lett*. **1978**, 3529.

(9) "A Simple Model for Predicting the Effect of Substituents on the Rates of Thermal Pericyclic Reactions," Carpenter, B.K. *Tetrahedron* **1978**, *34*, 1877.

**Brief biographies of guest editors**

Stephanie Hare is a computational chemist at Atomwise. She was a postdoc with Barry from September 2018 to April 2020, which generated one research paper. Despite their short time working together, Barry's influence has always been and continues to be very impactful to her research interests and goals.

Allan Pinhas is Professor of Chemistry at the University of Cincinnati. He was a graduate student at Cornell from 1975 to 1979 working with Barry. He was Barry's first Ph.D. student.

Julia Rehbein is a Professor of Organic Chemistry at the University of Regensburg, Germany. She worked with Barry during her postdoctoral stay in Cardiff, UK starting in 2009. She not only got inspired by his approach to see chemistry from a dynamic perspective, but also from his way of teaching. "Keep the momentum and keep Occam's razor sharp", is what she took from that great time with him!

Dean Tantillo is a Professor of Chemistry at UC Davis. He attended Barry's group meetings when he was a postdoc with Roald Hoffmann at Cornell and the three of them published one paper together. Barry's work has inspired Dean for many years and, aside from that of my PhD and postdoc mentors, has had the most significant influence on his career.

Stephen Wiggins is a Professor of Applied Mathematics at the University of Bristol. He has been collaborating with Barry since around 2012. They have written 9 papers together, and have collaborated in a large EPSRC grant during this time.